\documentclass{svjour3}
\usepackage{graphicx}

\begin{document}

\title{Localization of the relative phase {\it via} measurements\thanks{This work was supported by the Academy of Finland (Acad. Res. Fellowship 00857 and %%@
projects 7111994 and 7118122).
}}

\author{G.~S.~Paraoanu}

\institute{Low Temperature Laboratory, Helsinki University of Technology,\\
P.O. Box 5100, FIN-02015 TKK, Finland\\email:paraoanu@cc.hut.fi}

\date{\today}

\maketitle

\begin{abstract}
When two independently-prepared Bose-Einstein condensates are released from their corresponding traps,
the absorbtion image of the overlapping clouds presents an interference pattern. Here we analyze a
model introduced by Javanainen and Yoo \cite{javanainen}, who considered
two atomic condensates described by plane waves propagating in opposite directions.
We present an analytical argument for the measurement-induced breaking of the relative
phase symmetry in this system, demonstrating how the phase gets localized after a large enough number
of detection events.



\end{abstract}

\section{Introduction}

A long-standing problem in physics is to understand how relative phases
are established between relatively independent systems that have never been in contact
with each other. In Anderson's famous {\it gedankenexperiment} \cite{anderson} two buckets of superfluid Helium are suddenly
put in contact through a weak link: since there is no well-defined phase between the two condensates\footnote{It is important
to notice that one should first cool the systems into the superfluid phase and only then connect them through the weak link.
For otherwise, if the weak link is realized before, the system cools into the ground state, and a relative phase equal to zero
will be established between the two components of the order parameter.},
the question is then if a Josephson current will flow between them and why.
More recently, experiments done at MIT \cite{andrews} have established that clouds of Bose-Einstein
condensates exhibit an interference pattern when released from the traps and imaged using a standard CCD camera.
It soon became clear \cite{javanainen,ycastin}, that
for standard single-shot absorbtion measurements the data cannot be explained simply by taking the average of the
density operator; instead, what is in fact recorded are higher-order correlations.
In a landmark paper by Javanainen and Joo \cite{javanainen}, a detection experiment on a system consiting of two counterpropagating beams is simulated numerically;
it is found, surprisingly, that the result shows an interference pattern although the initial state is fragmented.
In another important paper \cite{ycastin}, a two modes/two detectors simplified model was introduced in analogy with the
Hanbury Brown and Twiss experiment from optics. The essential feature captured by this model is that
the detectors are such that they cannot distinguish between atoms prepared independently in two orthogonal modes.
Other authors have extended the investigation of the effects of measurement on various other types of
many-body fragmented states involving spins \cite{laloe} or rotating and attractive-interaction condensates \cite{erich}.
One important prediction of these very elegant theories
is that the interference fringes will be shifted randomly (corresponding to a random phase between the condensates)
from one run to the other of the experiment.

Although for the original MIT interference experiment \cite{andrews} it can be argued that the randomness in the %%@
position of the fringes could also be caused by fluctuations in the trapping Hamiltonian \cite{ketterleirreproducible}, a
latter generation of experiments succeeded in getting better control over the
external trapping parameters, as demonstrated by
the reproducibility of the results when phase states are prepared and measured.
Such interference experiments have  been done in optical lattices \cite{dalibard}
 and with on-chip splitters \cite{ketterlenew}. The result is that indeed
a random phase is detected for each realization of the experiment, leaving little doubt that
the interpretation \cite{javanainen,ycastin} is correct. Moreover, for some of these experiments the interaction energy can be %%@
neglected due to the low density of particles when the clouds corresponding to neighbouring sites overlap.

\section{Model}

\begin{figure}
\centering
\includegraphics[width=5in]{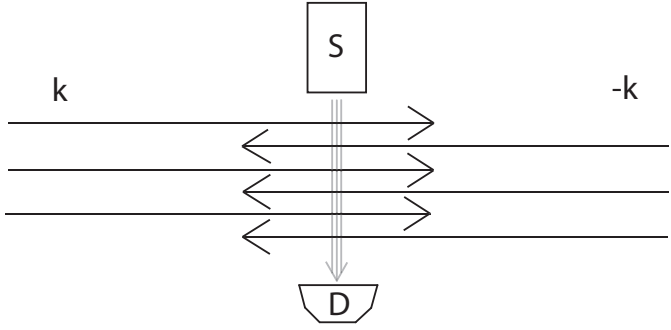}
\caption{Schematic of the experiment: atoms are released from left and right with wavevectors $k$ and $-k$ repectively and
a source S ({\it e.g.} a laser)  is used to take images of the overlapping region by sending photons to a detector D ({\it e.g.} a CCD camera).}
\label{setup}
\end{figure}

In a real experiment, such as those mentioned above \cite{andrews,dalibard,ketterlenew}, the atoms are
released from a potential that confines them in the ground state, with an order parameter given by the Gross-Pitaevskii
equation. In this initial state, the two clouds have mostly potential and interaction energy (as given for example by the
Thomas-Fermi approximation); but, after the confining potential is switched off, this energy is rapidly transformed
into kinetic energy. Depending on the particular setup, it can happen that at the time of the overlap of the two
expanding clouds ({\it i.e.} of their time-dependent Gross-Pitaevskii wavefunctions) the interaction energy is completely
negligible and therefore the atoms' motion is described by the standard Schr\"odinger equation. If this is not the case, the
problem of taking the interaction into account is a challenging one \cite{pra}.  In this paper, we will neglect the interaction
between atoms, and, to simplify the situation further \cite{javanainen} we will take the two atomic clouds as described by two plane waves
of momenta $\pm \hbar k$. A laser is then used to perform absorbtion (destructive) measurements, as in Fig. (\ref{setup}). We note here that treating the problem without this simplification is possible by generalizing the
results below, since approximate expressions for the order parameters of the expanding condensates are known.
We work with periodic boundary conditions over a length $L$; therefore
$k = (2\pi /L)l$, where $l$ is an integer number.

The initial state of the system is a fragmented or Fock state,
\begin{equation}
\mid \left(N/2\right)_{k},
\left( N/2\right)_{-k}\rangle =\frac{1}{(N/2)!}\hat{b}_{k}^{\dagger N/2}
\hat{b}_{-k}^{\dagger N/2}\mid 0 \rangle
.\label{st}
\end{equation}

The
Hamiltonian of the system has only the term corresponding to kinetic energy,
\begin{equation}
\hat{H} =
\frac{-\hbar^2}{2m}\int dx\hat{\psi}^{\dagger}(x)\nabla^{2}\hat{\psi}(x) = \frac{\hbar^2}{2m}\sum_{q}
\hat{b}^{+}_{q}\hat{b}_{q},
\end{equation}
so the atoms are not aware of each other's existence and
as a result the state of the system at a later time $t$, satisfying the Schr\"odinger equation
$i\hbar (d/dt)\mid \Psi
\rangle= H \mid \Psi
\rangle$,
 will be
\begin{equation}
\mid \Psi
\rangle =e^{-i\hbar k^{2}Nt/2m}\mid \left(N/2\right)_{k},
\left( N/2\right)_{-k}\rangle ,\label{st}
\end{equation}
where the phase factor is irrelevant. At the moment of detection, the state of the system is therefore a fragmented state.

The field operator can then be expanded in the corresponding mode operators,
\begin{equation}
\hat{\psi} (x) = \sum_{q}\frac{e^{iqx}}{\sqrt{L}}\hat{b}_{q},
\end{equation}
and, taking  into account that only  $k$ and $-k$ are occupied, we can truncate the expansion to
 $\hat{b}_{k}$ and
$\hat{b}_{-k}$ only,
$\hat{\psi} (x) \approx (1/\sqrt{L})(\hat{b}_{k}e^{ikx} + \hat{b}_{-k}e^{-ikx})$.

Fragmented states do not seem at first sight too exciting candidates for interference: consider
$\mid \Psi \rangle = \mid (N/2)_{k}, (N/2)_{-k} \rangle$ in the limit
of a large number of particles $N\gg 1$. Then
the probability of finding a particle at a point $x$ is the same for every point so
the average density does not present any spatial structure:
\begin{equation}
\langle \Psi \mid \hat{\psi}^{+} (x)\hat{\psi} (x) \mid \Psi \rangle = N/L.\label{dens}
\end{equation}
Contrast that with the case of a phase state,
\begin{equation}
\mid \varphi \rangle_{N} = \frac{1}{\sqrt{2^{N}N!}}(\hat{b}^{+}_{k}e^{i\varphi /2} +
\hat{b}^{+}_{-k}e^{-i\varphi /2})^{N}\mid 0 \rangle ,
\end{equation}
where the density
\begin{equation}
_{N}\langle \varphi \mid \hat{\psi}^{+} (x)\hat{\psi} (x)\mid \varphi \rangle_{N} = \frac{N}{L} (1+ \cos (2kx + \varphi )).\label{densphase}
\end{equation}
oscillates in space with a period given by the relative momentum $2\hbar k$ between atoms with wavevectors $\pm k$.
However, for a fragmented state something interesting seem to happen with the second order (density-density) correlation function,
\begin{equation}
\langle \Psi \mid \hat{\psi}^{+} (x_{1}) \hat{\psi}^{+} (x_{2})\hat{\psi}(x_{2})\hat{\psi}(x_{1}) \mid \Psi
\rangle = \frac{N^2}{L^2}(1+\frac{1}{2}\cos 2k(x_{1}-x_{2})), \label{cor}
\end{equation}
which reveals an interference structure. 

The same results can be obtained if the calculation is
performed in the phase representation: a fragmented state can be expanded in phase states
$
\mid \varphi \rangle_{N} = (2^{N}N!)^{-1/2}(\hat{b}^{+}_{k}e^{i\varphi /2} +
\hat{b}^{+}_{-k}e^{-i\varphi /2})^{N}\mid 0 \rangle ,
$
in the form \cite{ycastin}
\begin{equation}
\mid \Psi \rangle = \left( \frac{\pi N}{2}\right)^{1/4}\int_{0}^{2\pi}\frac{d\varphi }{2\pi}\mid \varphi \rangle_{N} .
\end{equation}
Now, using the orthogonality relation $_{N}\langle \varphi \mid \varphi '\rangle_{N} = 2\sqrt{2\pi /N}\delta (\varphi -\varphi ' )$,
we can easily recover the same expressions for the correlation functions Eqs. (\ref{dens},\ref{cor}) for the fragmented
states with the additional benefit of the insight that integration over phases washes out the interference fringes in the density
Eq. (\ref{densphase}) for phase states. What does this mean? The interference pattern in Eq. (\ref{cor}) is obtained by removing
an atom at one position and adding another one at a different position (correlations functions of this type do indeed test the
response of the system at one point given that an operator is applied at another point). This suggests that, in order to predict
the result of such an experiment, we have to carefully
consider the effect of the measurement process itself one atom at a time, which we do in the next section.

To make the connection with other insights into this problem, we first look at the combination of operators in Eq. (\ref{cor})
which is responsible for the interference term, and we find
the operator $\hat{A} = \hat{b}^{+}_{k}\hat{b}_{-k}$ introduced in \cite{anatoli}, which can be regarded
as the quantum observable giving the amplitude of the fringes;
the square amplitude of the interference
term in Eq. (\ref{cor}) is obtained from averaging the operator $\hat{A}^{+}\hat{A}=\hat{b}^{+}_{-k}\hat{b}^{+}_{k}\hat{b}_{k}\hat{b}_{-k}$,
namely
$\langle \Psi \mid \hat{b}^{+}_{-k}\hat{b}^{+}_{k}\hat{b}_{k}\hat{b}_{-k}\mid \Psi \rangle = (N/2)^2$.  Second, we notice that the two-detector
model presented in \cite{ycastin}
can be understood as a particular case of the model above, namely that in which detection can be realized only at positions
$x_{+}=0$ and $x_{-}=L/4l$
($kx_{-}=\pi /2$), leading to the "detector" operators $\sqrt{L}\hat{\psi}(x_{+})= \hat{b}_{k}+\hat{b}_{-k}$ and
$-i\sqrt{L}\hat{\psi}(x_{-})= \hat{b}_{k}-\hat{b}_{-k}$.

\section{Measurement}

What happens during the measurement of such a state? We divide the length $L$ into $M\gg q\geq 1$ intervals of length $L/M$ each,
and we consider $n$ atoms detected, $n_{1}$ atoms in the first interval $n_{2}$ atoms in the second ..., and $n_{M}$ in
the last interval. Following  the quantum-optics theory of detection, the many-body state after $n=n_{1}+n_{2}+ ....+n_{M}$
atoms are detected becomes
\begin{eqnarray}
&&\prod_{i=1}^{M}\hat{\psi} (x_{i})^{n_{i}}\mid \Psi \rangle_{N} = \nonumber \\
&& \left(\frac{2}{L}\right)^{n/2}\sqrt{\frac{N!}{n!}}\left(\frac{\pi N}{2}\right)^{1/4}\int_{0}^{2\pi}\frac{d\varphi}{2\pi}\prod_{i=1}^{M}\cos^{n_{i}}
(kx_{i}+\varphi /2)\mid \varphi \rangle_{N-n}.
\end{eqnarray}

The statement that we want to prove is that sequences of $n_{i}$ detections at positions $x_{i}$ occur with maximum probability if the
Born probability
rule $n_{i} = 2 (n/M)\cos^{2}(kx_{i}+\tilde{\varphi} /2)$, is satisfied, where $\tilde{\varphi}$ is
an arbitrary angle. We proceed by mathematical induction: given that this statement is true for $n$ detections, we want to prove that the next detection event
will tend to occur with maximum probability such that the law above is satisfied. As the number of detection events increases, the next
detection events tend to "reinforce" this law, in the sense that the probability for events that do not satisfy the Born probability distribution
becomes exponentially small. The tendency of phase localization increases exponentially from one detection to the other.

We start with the first two detection events and we go back to understanding the structure of Eq. (\ref{cor}).
The probability of detecting a particle at position $x_{i}$ with a system in a fragmented state is proportional to the first-order
correlation which is flat. That is, all points $x_{i}$ are equally probable, $P(x_{i}) = 1/M$.
Not so for the next detection event, for which the probability will be proportional to the second order correlation function,
\begin{equation}
P(x_{j}\mid x_{i})= \frac{1}{M}\left( 1+ \frac{1}{2}\cos (2kx_{j}-2kx_{i})\right) .
\end{equation}
Thus, a single detection at $x_{i}$ sets up a phase $2kx_{i}$ which will make the next detection event more probable at points
$x_{j}$ where $\cos (2kx_{j}-2kx_{i})=(1/2)(1+\cos^{2}(kx_{j}-kx_{i}))$
is large. This also means that after the first detection event the relative phase of the system is no longer undefined, which can be checked directly for example by
calculating the fluctuations of the phase operator \cite{me} in the state $\hat{\psi}(x_{i})\mid N/2, N/2\rangle$. A more elegant way is to employ
Bayes' theorem \cite{durr,zin} and attempt to calculate the conditional probability that the relative phase takes a certain value $\varphi$ given a
detection event at $x_{i}$, $P(\varphi \mid x_{i})$: the probability of obtaining a count at $x_{i}$ is $P(x_{i})=1/M$, and the probability of measuring a certain phase $\varphi$
on the fragmented state is also equally distributed, $P(\varphi ) = 1/2\pi$.
If the system would have been prepared in a phase state $\mid \varphi \rangle_{N}$, the probability to count a particle at
position $x_{i}$ had been
$P(x_{i}\mid \varphi) = M^{-1}[1+\cos (2kx_{i}+\varphi )]$. Bayes' theorem allow us to change our past description of  a system given that further information is available:
in our case, it relates the probabilities above as
\begin{equation}
P(\varphi \mid x_{i})=\frac{P(\varphi )}{P(x_{i})}P(x_{i}\mid \varphi) = \frac{1}{2\pi}[1+\cos (2kx_{i}+\varphi )],
\end{equation}
showing indeed that the most probable phases are those that get close to $2kx_{i}$.

We now look at what happens if the statement above is valid up to the $n$'th detection event.
This means that we identified an angle $\tilde{\varphi}_{n}$ so that
$2 (n/M)\cos^{2}(kx_{i}+\tilde{\varphi}_{n} /2)$
closely resembles the histogram ${n_{i}, x_{i}}$.

We first notice that this results in an extremum for the probability amplitude:
\begin{eqnarray}
&&0 = \frac{d}{d\varphi}\prod_{i=1}^{M}\cos^{n_{i}}(kx_{i}+\varphi /2) = \nonumber \\
&&-\sum_{i}n_{i}\sin (k_{i}+\varphi /2)
\cos^{n_{i}-1}(kx_{i}+\varphi /2)\prod_{j\neq i}^{M}\cos^{n_{j}}(kx_{j}+\varphi /2).  \label{porras}
\end{eqnarray}
It is immediate to see that this equation is satisfied
for $n_{i}= 2 (n/M)\cos^{2}(kx_{i}+\tilde{\varphi}_{n} /2)$; indeed
plugging this expression
into Eq. (\ref{porras}) results in
\begin{eqnarray}
&& \frac{d}{d\varphi}\prod_{i=1}^{M}\cos^{n_{i}}(kx_{i}+\varphi /2) \mid_{\varphi = \tilde{\varphi}_{n}}= \nonumber \\
&&-\frac{n}{M}\left(\sum_{i=1}^{M}\sin (2k_{i}x_{i}+\tilde{\varphi}_{n})\right)\times
\prod_{i=1}^{M}\cos^{n_{i}}(kx_{i}+\tilde{\varphi}_{n}/2) = 0,
\end{eqnarray}
since the first sum vanishes.
We note that our solution gives the correct normalization condition  $\sum_{i=1}^{M}n_{i}=n$;
and that
$n_{i}/n=(L/M) (\sqrt{2/L}\cos (kx_{i}+\tilde{\varphi}_{n} /2))^2$ is precisely Born's probability law corresponding to a wavefunction
$\sqrt{2/L}\cos(kx_{i}+\tilde{\varphi}_{n} /2)$. After $n$ detections the amplitude becomes peaked around a certain value
$\tilde{\varphi}_{n}$ of the relative phase,
which satisfies as closely $n_{i}=2(n/M)\cos^{2}(kx_{i}+\tilde{\varphi}_{n} /2)$; further absorbtion events will make this amplitude
even more peaked. The final result is that after a large enough number of detection events, the initial fragmented
wavefunction collapses onto one of the phase states. (This result is rather general; it does not depend on the specific form
of the wavefunction, as long as it is normalized \cite{pra}.)
To convince ourselves that this is indeed the case,
consider the square of the amplitude probability
\begin{equation}
f(\varphi ) = \prod_{i=1}^{M}\cos^{2n_{i}}(kx_{i}+\tilde{\varphi}_{n} /2).
\end{equation}
Since by definition this quantity is positive, we can take the $\ln$ and expand it in a Taylor series. As expected, the first
order derrivative is zero at $\varphi = \tilde{\varphi}$, indicating an extremum,
\begin{equation}
\frac{d}{d\varphi} \ln f(\varphi )\mid_{\varphi = \tilde{\varphi}_{n}} =
-\frac{n}{M}\sum_{i=1}^{M}\sin (2kx_{i}+\frac{\tilde{\varphi}_{n}}{2}) = 0,
\end{equation}
while in the second order we get
\begin{equation}
\frac{d^2}{d\varphi^2} \ln f(\varphi )\mid_{\varphi = \tilde{\varphi}_{n}} = -n.
\end{equation}
This means that we have a indeed a maximum at $\tilde{\varphi}$ and
\begin{equation}
f(\varphi ) = f(\tilde{\varphi})e^{-\frac{1}{2}n(\varphi - \tilde{\varphi}_{n})^{2}}.
\end{equation}

The essential thing to notice is that the variance of this normal distribution function is simply $n^{-1}$,
increasing with the number of detection events but independent on any particular outcome (i.e. how many detection events
$n_{i}$ we had at position $x_{i}$). After only a few detection events this will result in a sharply peaked distribution around
the value $\tilde{\varphi}$ (which depends on the distribution $\{ n_{i}, x_{i} \}$). Indeed, it has been demonstrated
experimentally that extracting a small fraction of the total number of atoms is enough to establish a relative phase \cite{pritchard}.
Assuming $N-n>>1$, and using the
quasi-orthonormalization relation for the phase states, we obtain the normalized wavefunction in the form
\begin{equation}
\mid\Psi^{(n)}\rangle = \frac{Nn}{4\pi}\int_{0}^{2\pi}
d\varphi e^{-\frac{1}{4}n(\varphi - \tilde{\varphi}_{n})^{2}}\mid \varphi \rangle_{N-n}.\label{hir}
\end{equation}

We now take the n-times collapsed wavefunction $\mid \Psi ^{(n)}\rangle$ resulting after a sequence of $\{x_{j},n_{j}\}$ detections
and ask what is the probability of detecting a particle at a position $x$. This quantity will
be proportional with $\langle \Psi ^{(n)} \mid \hat{\psi}^{+}(x_{i})\hat{\psi}(x_{i})\mid\Psi ^{(n)} \rangle$.
Using the properties of Gaussian integrals, one can easily
calculate
\begin{equation}
\langle \Psi ^{(n)} \mid \hat{\psi}^{+}(x_{i})\hat{\psi}(x_{i}) \mid\Psi ^{(n)}\rangle =
\sqrt{\frac{4\pi N (N-n)}{L}}\left[1+ e^{-1/2n}\cos(2kx_{i}+\tilde{\varphi})\right].
\end{equation}
resulting in a  conditional probability
\begin{equation}
P(x_{i}\mid \{x_{j},n_{j}\}) = \frac{1}{M}\left[1+ e^{-1/2n}\cos(2kx_{i}+\tilde{\varphi}_{n})\right], \label{uuuhh}
\end{equation}
(or, equivalently, a conditional probability density $L^{-1}\left[1+ e^{-1/2n}\cos(2kx_{i}+\tilde{\varphi})\right]$).
From this expression it follows
that events which fall far from the Born probability distribution $[\sqrt{2/L}\cos(kx_{i}+\tilde{\varphi}_{n} /2)]^2$ are very unlikely to occur.
Consider for example a detection at a minimum of this probability, say $x_{i}=(\pi-\tilde{\varphi}_{n})/2k$; the probability
of this event is
\begin{equation}
P((\pi-\tilde{\varphi}_{n})/2k\mid \{x_{j},n_{j}\}) = \frac{1}{M}(1-e^{-1/2n})\stackrel{n\gg1}{\rightarrow}0 .
\end{equation}
Another way to look at this effect is to think in terms of probabilities of detecting a phase $\varphi$: from Eq. (\ref{hir}) above, it is obvious that
the probability of detecting a phase $\varphi$ is a Gaussian centered at $\tilde{\varphi}_{n}$ and of variance $1/\sqrt{n}$,
\begin{equation}
P(\varphi ) = \sqrt{\frac{2n}{\pi}}e^{-n(\varphi -\tilde{\varphi}_{n})^2}.
\end{equation}
We now try to get an estimate for the behavior of the optimal phase $\tilde{\varphi}$ as the number of detection events increases.
To do we that we evaluate
the difference between the optimal phases for two consecutive detection events, $\tilde{\varphi}_{n+1} - \tilde{\varphi}_{n}$.
Suppose that a detection event occurs at $x_{i}$; then we find
the correction made to the optimal value $\tilde{\varphi}$ by applying
the operator $\hat{\psi}^{+}(x_{i})$ to the state $\mid \psi^{(n)}\rangle$ and, as before, finding the extremum of the quantity
$\cos (kx_{i}+\varphi/ 2)\exp [-n(\varphi - \tilde{\varphi}_{n})/2]$ appearing under
integration over $\varphi$. We have also checked numerically that for $n\gg 1$ this quantity has a unique,
sharp maximum value in the interval $[0,2\pi]$, which is a solution of the equation
\begin{equation}
\tilde{\varphi}_{n+1}-\tilde{\varphi}_{n} = -\frac{1}{n}\tan\left( kx_{i}+\frac{\tilde{\varphi}_{n+1}}{2}\right). \label{hoo}
\end{equation}
An analysis of this expression shows that the new value $\tilde{\varphi}_{n+1}$ is within $1/n$ of the old value $\tilde{\varphi}_{n}$ for events that
have a finite probability (i.e. for which $[\sqrt{2/L}\cos(kx_{i}+\varphi /2)]^2$ is not too close to zero). This result is intuitively clear: if a
detection event occurs at a maximum of $[\sqrt{2/L}\cos(kx_{i}+\varphi /2)]^2$, the phase is not modified much. More change could happen if some improbable event
($[\sqrt{2/L}\cos(kx_{i}+\varphi /2)]^2$ small) occurs, and this could produce significant changes in the phase  only if
$n$ is not too large
(in other words, the value of the phase is decided in the beginning: the more detection events accumulate, the less is the
influence of an improbable event, even if it occurs).
As a result, the evolution of the phase $\tilde{\varphi}_{n}$ is given by a discrete stochastic process,
with steps typically of the order of $1/n$ given by Eq. (\ref{hoo})
and probabilities Eq. (\ref{uuuhh}), leading very fast to localization.

In the case of the particular model of  two-modes/two-detectors \cite{ycastin} we obtain for the conditional probabilities  Eq. (\ref{uuuhh})
that an atom is counted
in the "+" or "-" detector given a previous sequence of $n_{+},n_{-}$ detections
($n=n_{+}+n_{-}$) as
\begin{equation}
P(\pm \mid n_{+},n_{-}) = \frac{1}{2}(1\pm e^{-1/2n}),
\end{equation}
and the equivalent of the equation Eq. (\ref{hoo}) is
\begin{equation}
\tilde{\varphi}_{n+1}-\tilde{\varphi}_{n} = \mp\frac{1}{n}\left[\tan(\tilde{\varphi}_{n+1}/2)\right]^{\pm 1}.
\end{equation}
The same phenomenon occurs here as well for relatively large $n\gg 1$: the next value $\varphi_{n+1}$ will fall within $1/n$
of the previous value, approximating better and better the final value of the phase.
To give a numerical example, let us suppose that at $n=500$ we have obtained $\tilde{\varphi}_{500}=0$. A detection in the $+$ channel will only "confirm" the value
zero for the relative phase, since the equation $\tilde{\varphi}_{501}= -2\times 10^{-3}\times \tan(\tilde{\varphi}_{501}/2)$
has $\tilde{\varphi}_{501}=0$ as solution. This value can be changed though by a detection in the other channel: indeed, solving numerically the equation
$1=500\times \tilde{\varphi}_{501}\times\tan(\tilde{\varphi}_{501}/2)$ results in $\varphi_{501}=0.063$, a small correction
but worth considering. However, the probability for a detection in the - channel is only $P(-)= 5\times 10^{-4}$. Clearly, to get
a significant change in the value of the relative phase one would need a series of such events, which is very improbable.

\section{Conclusions}
We have presented an argument based on mathematical induction for the localization of phase under absorbtion
measurement of a Fock state consisting of counterpropagating atomic plane waves. We show that, as the number of atoms
detected increases, an arbitary phase builds up.

\section{Acknowledgments}
The author wishes to thank A. Polkovnikov for stimulating discussions.

\end{document}